\documentstyle[aps,epsf,multicol,prl]{revtex}
\begin{document}
\draft

\title{Andreev current in finite sized carbon nanotubes}

\author{Smitha Vishveshwara$^1$, Cristina Bena$^1$, 
Leon Balents$^1$, Matthew P. A. Fisher$^2$}
\address{$^1$Department of Physics, University of California, 
Santa Barbara, CA 93106 \\
$^2$Institute for Theoretical Physics, University of California,
Santa Barbara, CA 93106--4030
}
\date{\today}
\maketitle

\begin{abstract}
We investigate the effect of interactions on Andreev current
at a normal-superconducting junction when the normal phase
is a Luttinger liquid with repulsive interactions.
In particular, we study the system of 
a finite sized carbon nanotube 
placed between one metallic and one superconducting lead. 
We show that interactions and finite size effects give rise to significant 
deviations from the standard picture of Andreev current at a normal-
superconductor junction in the nearly perfect Andreev limit.
\end{abstract}


\begin{multicols}{2}

In recent years, the behavior of superconductors (SC)
in contact with Luttinger liquids (LLs) has
commanded attention in both theory \cite{Jos2,Affleck} and experiment
\cite{Jos1,Morpurgo}. Josephson
junctions made by sandwiching a Luttinger liquid between two superconductors
have led to intriguing results such as 
critical currents orders of magnitude 
larger than expected \cite{Jos1}.
Experimental 
study of Andreev physics at a niobium superconductor -- carbon
nanotube junction \cite{Morpurgo} has yielded 
significant deviation at low temperatures
from the standard picture of Andreev current in a non-interacting
one-dimensional electron gas -- superconductor
junction \cite{BTK}. 
As it has been  predicted \cite{BFK} and 
shown \cite{nanexp} that single-walled metallic carbon nanotubes 
(NT) exhibit Luttinger liquid behavior, 
systematic analyses of any set-up involving their electronic properties would 
require taking into account the effect of interactions.

Here we study the Andreev physics
in a SC-NT-metallic lead junction, focusing on the effects of the
strong repulsive interactions and of the finite size of the nanotube. 
We focus on energy scales well below the energy gap $\Delta$ 
of the superconductor. Thus, throughout the energy range
of interest (i.e for all the values
of temperature or of the applied voltages we consider), 
the only excitations allowed to exit or enter
the superconductor are Cooper pairs and not single electrons. 
In particular we focus on the limit of almost perfect Andreev reflection
at the SC-NT interface (i.e. very low normal backscattering). We also assume 
that the nanotube-metal contact is ideal and that the nanotube
continues adiabatically into the metallic lead. 
Under these assumptions we study how a small amount of 
backscattering at the SC-NT interface would influence
 the electrical properties
of the junction, in particular the behavior of the conductance.

The treatment we use to obtain the value of the current as a function
of the applied voltage is a non-equilibrium Keldysh technique,
perturbative in the bare backscattering strength '$u$' \cite{Keldysh,KF}. 
Characteristic of Luttinger liquids,
the amount of backscattering can strongly increase when
the energy at which the system is probed decreases. 
Hence, perturbation theory holds good only above an energy scale $E_c \approx 
\epsilon_0 (u/\epsilon_0)^{2/(1-g)}$,
where $g$ measures the interaction strength
( $g=1$ in the absence of interactions). For metallic nanotubes, 
 $\epsilon_0 \approx 1eV$ is the sub-band spacing \cite{BFK}, and 
$g \approx 0.25$ \cite{BFK,nanexp},
corresponding to strong repulsive interactions.
In the set-up considered here, the effect of the finite length $L$ of the
nanotube becomes important below the finite size energy scale  $\hbar v/L$.
Here, $v=v_F/g$
is the velocity of the charge-carrying quasiparticles in the nanotube, 
where $v_F \approx 10^6 m/s$ is the Fermi velocity.
Effects of finite size 
can be captured in the perturbative approach, as done here,
provided $\hbar v/L \gg E_c$.

To summarize our results, a numerical analysis reveals 
that at zero temperature, 
the conductance shows a marked drop with decreasing voltage
as a consequence of 
LL physics, consistent with renormalization group arguments 
similar to the ones derived in \cite{Affleck}.
At voltages much smaller than the finite size energy $\hbar v/L$, 
the conductance levels off to a constant.
In addition, it exhibits small spikes with a voltage
spacing of $ \pi \hbar v/2 L$ 
(about $2-3 meV$ for a nanotube of micron length), reminiscent of resonance
peaks from quasi-bound states for charge carriers confined within the length 
of the tube.

We now present the explicit calculation 
yielding the conductance as a function of applied voltage
for the SC-NT-metal system described above. 
The s-wave SC 
lies in the region $x<0$ and we assume it to be ideally 
contacted to a finite size nanotube of
length $L$ in the region $0<x<L$ which continues adiabatically 
into a metallic lead for $x>L$.
We model the system in the semi-infinite region $x>0$, as a four channel LL
with interaction parameters appropriate for the nanotube up to
$x=L$, and appropriate for no interactions
for $x>L$.
The bosonized Hamiltonian for this system in the absence of
any normal backscattering is given by
\begin{equation}
H_0 = \int_0^{\infty} dx \sum_{a} 
v_a(x)[{1 \over {g_a(x)}}(\partial_x\theta_a)^2 + g_a(x) 
(\partial_x \phi_a)^2].
\label{hn}
\end{equation}
For simplicity, we have set the constants $\hbar=e=k_B=1$.
Note that $a=\rho_{\pm}, \sigma_{\pm}$ correspond to the four free sectors 
of the theory and are obtained by linear transformations 
from the spin-channel indices ($1\!\uparrow,1\!\downarrow, 
2\!\uparrow,2\!\downarrow$)\cite{BFK}.
The relation between the bosonic fields 
$\theta_{i \alpha}$,
$\phi_{i \alpha}$ ($i=1/2, \alpha=\uparrow/\downarrow$) and
the original chiral right-/left-moving 
electron fields $\Psi_{i R/L \alpha}$ 
is expressed through
the bosonization procedure via the transformation
$\Psi_{i R/L \alpha} \sim e^{i(\phi_{i \alpha} \pm \theta_{i \alpha})}$.
In the nanotube region $0<x<L$ where interactions are present 
in the net charge density $\rho+$, we take 
$g_{\rho+}(x) =g \approx .25$, and $g_{\rho_-,\sigma_{\pm}}=1$. Also,
in the non-interacting region $x>L$, we take  
$g_a(x) =1$ for all $a$'s. The velocities of the free modes
are given by $v_a(x)=v_F/g_a(x)$. 
The total charge density is $\rho_{tot}=2 
\partial_x\theta_{\rho_+}/\pi$. 

In the almost perfect Andreev limit,
the electrons incident from the nanotube side on the SC-NT interface
reflect back as holes with opposite spin: 
$\Psi_{i L \uparrow/\downarrow}(0)=
\Psi_{i R \downarrow/\uparrow}^{\dagger}(0)$ and
$\Psi_{i R \uparrow/\downarrow}(0)=
\Psi_{i L \downarrow/\uparrow}^{\dagger}(0)$, where $i$ refers to
the channel indices $1$ and $2$. In the bosonized language,
these boundary conditions become
$\phi_{\rho_{\pm}}(0)=0$ and $\theta_{\sigma_{\pm}}(0)=0$.

The weak normal backscattering at the SC-NT junction 
can be modeled by modifying the Hamiltonian to $H=H_0+H'$, with:
\begin{eqnarray}
H'&&={u \over 8}\sum_{i=1,2} \sum_{\alpha=\uparrow,\downarrow}
[\Psi_{i R \alpha}^{\dagger}(0) \Psi_{i L \alpha}(0)
+h.c.]\nonumber \\&& = u \cos[\theta_{\rho_+}(0)]
\cos[\theta_{\rho_-}(0)],
\label{backsc}
\end{eqnarray}
where the bosonized form takes into account
the Andreev boundary conditions at the SC-NT interface.
For simplicity, we choose not to include the backscattering processes 
where particles can flip their band index
since these terms do not give rise to any new physics.

Following Ref.\cite{KF}, we integrate out the $\phi$ variables in the
action, as well as the entire $x$-dependence away from
$x=0$. The resulting unperturbed imaginary time Euclidean action becomes
\begin{eqnarray}
S_0^{E}=\frac{1}{\beta}\sum_n {{|\omega_n|} \over {\tilde{g}(\omega_n)}} 
|\theta_{\rho_+}(\omega_n)|^2+\frac{1}{\beta}\sum_n |\omega_n| 
|\theta_{\rho_-}(\omega_n)|^2,
\label{s0}
\end{eqnarray}
where $\beta$ is inverse temperature. Here the imaginary time
Fourier transforms for all fields '$A$' are defined in the standard
fashion $A(\omega_n)=\int_0^{\beta} d\tau'
A(\tau') e^{i \omega_n \tau'}$, $\omega_n
=2 \pi n/\beta$.
The spatial variations of the interaction 
parameter $g(x)$ and of the velocity 
$v(x)$ are reflected by the fact that the 
effective interaction parameter $\tilde{g}(\omega_n)$ 
is frequency dependent and has the form:
\begin{eqnarray}
\tilde{g}(\omega_n)={{g(1-y)^2} \over {1-4 g y -y^2}} \text{,    with   }
y=\Big(\frac{1-g}{1+g}\Big)e^{-|\omega_n \tau|}.
\label{geff}
\end{eqnarray}
Here $\tau=2 L/v$ is the time it takes a charge-carrying
quasiparticle with velocity $v$
to bounce back and forth between the ends of the tube.
The limits $L, \omega_n \rightarrow 0$ and $L,\omega_n \rightarrow \infty$
retrieve the expected form $\tilde{g}(\omega_n)=1$ and
$\tilde{g}(\omega_n)=g$ for a semi-infinite Fermi liquid and a 
semi-infinite nanotube respectively \cite{FG}.

Along the lines of Ref.\cite{KF}, we proceed to construct a 
real time Keldysh  action. We introduce $\theta_a^{\pm}$ fields
running over forward and backward paths in time.
We define
$\theta_a=(\theta_a^{+}+\theta_a^{-})/2$ and 
$\tilde{\theta}_a=\theta_a^{+}-\theta_a^{-}$. 
The resulting action is  $S=S_0+S_1+S_2$ where $S_0$ is the 
unperturbed action, $S_1$
describes the effect of the weak backscattering at the SC-NT
interface, and $S_2$ captures the effect of applying a
chemical potential difference $V=\partial_t a$. Thus
\\
\phantom{1}
\put(-270,-5){\line(1,0){255}}
\put(-15,-5){\line(0,1){10}}
\end{multicols} 
\begin{eqnarray}
S_0=&&\int{d \omega \over {2 \pi}} {{\omega} \over 2}
[{1\over {g(\omega)}}+{1\over {g(-\omega)}}]\coth{{\omega} \over
{2 T}} |\tilde{\theta}_{\rho+}(\omega)|^2
+\int{d \omega \over {\pi}} {{\omega} \over 
{g(\omega)}} \tilde{\theta}_{\rho+}(\omega) \theta_{\rho+}(-\omega)
\nonumber \\&&
+\int{d \omega \over {2 \pi}} {\omega}
\coth{{\omega} \over
{2 T}} |\tilde{\theta}_{\rho-}(\omega)|^2
+\int{d \omega \over {\pi}} {\omega} 
\tilde{\theta}_{\rho-}(\omega) \theta_{\rho-}(-\omega),
\nonumber \\
S_1=&&i u
\int dt \{\cos[\theta_{\rho_+}^+(t)] \cos[\theta_{\rho_-}^+(t)]-
\cos[\theta_{\rho_+}^-(t)]\cos[\theta_{\rho_-}^-(t)]
\},
\nonumber \\
S_2=&&\int d\omega {{2 \omega} \over {\pi}}[a(\omega) 
\tilde{\theta}_{\rho+}(-\omega)+\eta(\omega)\theta_{\rho+}(-\omega)],
\label{action}
\end{eqnarray}
\begin{multicols}{2}
where
$g(\omega)$ is the analytically continued version of $\tilde{g}
(\omega_n)$ in Eq.(\ref{geff}) with
$|\omega_n \tau|$ replaced by $i \omega \tau$. For all fields '$A$', 
we have used the real time Fourier transform convention 
$A(\omega)=\int dt A(t) e^{i \omega t}$.
The source field $\eta$ allows for calculation of the current, 
$I(t)=  2 \dot{\theta}_{\rho+}(t)/\pi=(i/2\pi) ( \delta S /\delta \eta(t) 
|_{\eta=0})$. Average quantities may be derived
by taking expectation values with respect to 
the  Keldysh generating
functional, $Z=\int {\cal D}[\theta^+]{\cal D}[\theta^-]e^{-S}$.

Using the above Keldysh action, and treating the backscattering to 
lowest non-vanishing order in perturbation, we find 
the expectation value of the current to be
\begin{eqnarray}
I=8 {{e^2} \over h} V-I_B,
\label{current1}
\end{eqnarray}
where from here on, we reinsert factors of $e, \hbar$ and $k_B$.
The first term is associated with the  constant conductance 
$G_0=8 {{e^2} \over h}$ in the absence of backscattering.
As expected, this ideal Andreev conductance of the finite size nanotube
in the presence of a metallic lead is that of a four mode non-interacting
1D electron gas \cite{Jos2,FG}.
The backscattering current $I_B$ takes the form:
\begin{eqnarray}
I_B=\frac{e}{2 \pi} \Big(\frac{u}{\hbar}\Big)^2\!\int_0^{\infty}\!dt 
\sin\bigg[\frac{2 e V t}{\hbar}\bigg] e^{[\tilde{C}(t)-\tilde{C}(0)]} 
\sin[\tilde{R}(t)].
\label{current2}
\end{eqnarray}
In the above equation, $\tilde{C}(t)=
\sum_{a=\rho_{\pm}}C_a(t)$ and $\tilde{R}(t)=
\sum_{a=\rho_{\pm}} [R_a(-t)-R_a(t)]/2$.  
For each mode, $C_a(t)$ and $R_a(t)$ are 
the correlation and response functions respectively, with 
$C_a(t) = \langle\theta_a(t)\theta_a(0)\rangle_0$ and 
$R_a(t) = 
-i \langle\theta_a(t)\tilde{\theta}_a(0) \rangle_0$.
Their Fourier transforms are related by the fluctuation-dissipation
theorem, $C_a(\omega) = -\coth \big(\frac{\hbar\omega}{2k_B T}\big) 
Im[R_a(\omega)]$.
Here averages are with respect to the unperturbed action, and we have
$R_{\rho+}(\omega)=-i \pi g(\omega)/\omega$ and 
$R_{\rho-}(\omega)=-i \pi/\omega$

We now make a series expansion of
$\tilde{g}(\omega_n)$ in Eq.(\ref{geff}), $\tilde{g}(\omega_n)=g \sum_n
\alpha_n y^n$. When analytically continued, this gives:
\begin{eqnarray}
g(\omega)=g \sum_{n=0}^{\infty} \alpha_n 
\big(\frac{1-g}{1+g}\big)^n e^{i n \omega \tau}.
\label{expand}
\end{eqnarray}
Substituting the above in Eq.(\ref{current2}) and taking the derivative
with respect to the applied voltage gives the following reduction in the
conductance due to backscattering $G_B = \frac{d I_B}{d V}$,
in the limit $T \rightarrow 0$:
\\
\phantom{1}
\put(-270,-10){\line(1,0){255}}
\put(-15,-10){\line(0,1){10}}
\end{multicols}
\begin{eqnarray}
G_B=2 \frac{e^2}{h} \Big(\frac{u \tau}{\hbar}\Big)^2 
\sum_{k=0}^{\infty} \sin(\pi \sum_{n=0}^k\beta_n)
\int_{k}^{(k+1)} dx x 
\frac{\cos[2 e V x \tau /\hbar]}{[1+(\epsilon_0 x \tau/\hbar)^2]^{\beta_0}} 
\prod_{n=1}^{\infty} \Big|\Big(\frac{x}{n}\Big)^2-1\Big|^{-\beta_n},
\label{current3}
\end{eqnarray}
\begin{multicols}{2}
where the coefficients $\beta_n = g (\alpha_n/2) 
[(1-g)/(1+g)]^n$, for $n>0$, and $\beta_0=(1+g)/2$.
We have used the high energy cut-off $\epsilon_0$
to evaluate the Fourier transforms $C_a(t)$ and $R_a(t)$.
Here the terms involving $n \tau$ correspond
to physical processes of $n$ bounces of the quasiparticles 
at the boundaries of the nanotube.
Note that besides the weak backscattering at the 
SC-NT interface, the quasiparticles can also backscatter
at the nanotube-metallic lead junction even in the 
absence of a barrier, solely as a result of 
the mismatch of the values of the net charge and velocity 
of the free modes in the nanotube and in the metal.
\narrowtext
\begin{figure}
\epsfxsize=3.5in
\centerline{\epsffile{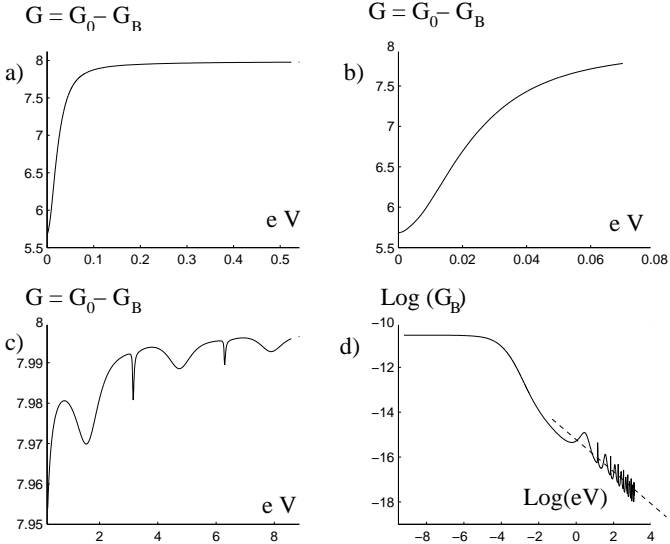}}
\vspace{.15truein}
\caption{The net differential conductance $G=dI/dV$ of the nanotube
set-up in units of $e^2/h$ as a function of 
applied voltage $e V$ in units of $\hbar/\tau$.
The values of the parameters 
are $g=.25$, $u=300 (\hbar/\tau)$ and $\epsilon_0=10^5 (\hbar/\tau)$.
a), b) and c) show the conductance for different ranges of applied
voltage. The log-log plot of $G_B$ in d) shows a constant at low voltages
and a power law $G_B\sim V^{g-1}$ on average for high voltages, as indicated by
the dashed line.}
\label{figcurrent1}
\end{figure}
The most revealing analysis of Eq.(\ref{current3})
comes from numerical evaluation. 
We restrict the infinite sums of
Eq.(\ref{current3}) to a finite number of terms, 
and introduce an explicit high energy cutoff $\epsilon_0$, 
in order to regulate singularities.
We ensure that errors coming from both truncations are negligible. 
In Fig.1, we plot the net differential conductance 
$\frac{d I}{d V}= G_0-G_B$, as the one of experimental
relevance. The conductance drops with decreasing voltage
 and levels off at voltages much smaller than $\hbar/\tau$.
To see why this might be expected, notice that 
at large voltages $eV \gg \hbar/\tau$, the
time scale at which the system is probed is much shorter than $\tau$,
and the conductance roughly behaves as if the nanotube were semi-infinite.
Characteristic of Luttinger liquid physics, it thus drops as $G_B \propto
u^2 (2 g e V/\epsilon_0)^{g-1}$ on average, as shown in Fig. 1 d). This 
limiting behavior can be seen directly in Eq.(\ref{current3}) by taking 
$\tau \rightarrow \infty$.
At low voltages $eV \ll \hbar/\tau$, the associated time scale is long enough 
to capture the effect of the metallic lead and of multiple backscattering
events at its interface. With decreasing voltage, the conductance 
ultimately levels off to a constant, as per Ohm's law for the metallic lead.
This limit can be obtained in Eq.(\ref{current3}) by setting
$\tau = 0$.

A striking feature of the plots is the presence of spikes 
at probe values
$e V = n\hbar\pi/\tau$, with '$n$' being an integer. As their
magnitude is minuscule compared to the net variation in 
conductance, it would be difficult to measure them in 
experiment. However, these resonances do exist, and are
signatures of the quasi-bound states that one would 
expect within the nanotube region, given that here the 
interaction parameter and velocity are different from those of the
metallic lead.

As a variation of the above set-up, let us now replace the nanotube
by a finite sized Luttinger liquid with only two transport 
channels (spin $\uparrow/\downarrow$). Such a situation can be realized
by using, for instance, an etched quantum wire.
The corresponding free modes carry net charge 
$\rho$ and spin $\sigma$, and are linear combinations of the two 
spin species. Andreev boundary conditions at
the superconducting junction require $\phi_{\rho}(0)=0, 
\theta_{\sigma}(0)=0$ in corresponding bosonized variables.
Thus, the system can be effectively described by a single channel in the
$\theta$ variables. This allows for us to study the particular situation where
the velocity of the charge mode in the Luttinger liquid $v=v_F/g$ would
equal the Fermi velocity $v_F^l$ in the metallic lead, i.e $v_F^l=v_F/g$.  
Hence, we can focus on the physics arising purely from the mismatch
of the charge of the elementary excitations in the Luttinger liquid and the 
lead. This would not have been possible for the case of the nanotube
as it is described by two modes moving at different velocities, and 
matching the velocity of one mode to the Fermi velocity of the metallic lead
would cause a velocity mismatch in the other mode.
We calculate the conductance as a function of applied voltage
for this system  in a manner completely
analogous to the one described above for the nanotube. The major difference 
here is that we have only one mode with 
effective interaction parameter
\begin{equation}
\tilde{g}(\omega_n)= \frac{g}{1-(1-g)\exp(-|\omega_n\tau|)}.
\label{geffLL}
\end{equation}

The resulting conductance is plotted in Fig. 2. 
The magnitude of the resonances at 
$e V = n\hbar\pi/\tau$ spans a larger fraction of the net
variation in conductance compared to the case of the nanotube.
It is noteworthy that these resonances exist in spite of the fact that 
there is no mismatch of velocities of the free phonon modes in the Luttinger liquid
and in the metal. As expected, the phonons rebound at the Luttinger 
liquid-metal interface
solely due to the impedance mismatch in the charge sector.

\begin{figure}
\epsfxsize=2in
\centerline{\epsffile{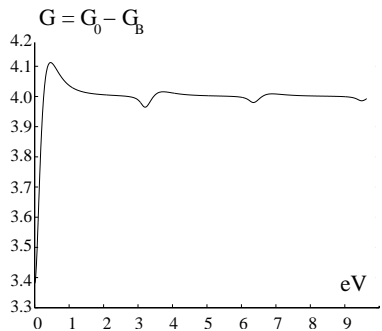}}
\vspace{0.15in}
\caption{The net conductance $G$ of the two-mode Luttinger liquid 
set-up in units of $e^2/h$ as a function of 
applied voltage $e V$ in units of $\hbar/\tau$.
The values of the parameters 
are $g=.25$, $u=1.4 (\hbar/\tau)$ and $\epsilon_0=10^4 (\hbar/\tau)$.}
\label{lutt}
\end{figure}
To summarize, we have looked at how the standard picture for Andreev
current through a superconductor-metallic wire junction gets
altered in the Andreev limit in the presence of interaction and finite size
effects. The Andreev conductance shows a reduction with decreasing voltage
which finally levels off at the lowest voltages. Finite size
effects also give rise to resonances manifest as small spikes in
the conductance.

Finally, turning to experiment, while Luttinger liquid behavior 
in nanotubes contacted to normal leads has been analyzed in great 
detail \cite{nanexp}, by no means has it been studied systematically 
in the presence of superconducting leads.
As seen here, one would
certainly expect Luttinger liquid effects to yield significant deviations
from the standard picture of Andreev physics for non-interacting
one-dimensional wires. Consistent with our assumptions, such experiments 
can be performed, for example, in  superconductor-nanotube junctions, in which
the superconducting gap energy is of the order of several $meV$, 
while for a nanotube of a few
microns, the finite size energy is in the range of a $meV$. 
At temperatures of the  order of $100mK$, thermal effects
are expected to be negligible. 
These conditions are well within experimental reach, and systematic
analyses of such set-ups could potentially reveal rich physics arising from
bringing Luttinger liquids in contact with superconductors.

This work was supported by NSF grants DMR-9985255, DMR-97-04005,
DMR95-28578, PHY94-07194, and the Sloan and Packard foundations.

\end{multicols}

\begin{references}
\bibitem{Jos2}
R. Fazio, F. W. J. Hekking
and A. A. Odintsov, Phys. Rev. Lett {\bf 74}, 1843 (1995);
D. L. Maslov, M. Stone, P. M. Goldbart and
D. Loss, Phys. Rev. B {\bf 53}, 1548 (1996);
Y. Takane, J. Phys. Soc. Japan, {\bf 66}, 537 (1997); 
R. Fazio, F. W. J. Hekking, A. A. Odintsov and R. Raimondi, cond-mat/9811217.
\bibitem{Affleck} I. Affleck, J.-S. Caux and A. M. Zagoskin, 
Phys. Rev. B {\bf 62}, 1433 (2000).
\bibitem{Jos1} A. Yu. Kasumov et al., Science {\bf 284}, 1508 (1999).
\bibitem{Morpurgo} A. F. Morpurgo et al., Science {\bf 286}, 263 (1999).
\bibitem{BTK} G. E. Blonder, M. Tinkham, and T. M. Klapwijk, Phys. Rev. B
{\bf 25}, 4515 (1982,.
\bibitem{BFK} C. L. Kane, L. Balents and M. P. A. Fisher, Phys. Rev. Lett. 
{\bf 79}, 5086 (1997); R. Egger and A. Gogolin, Phys. Rev. Lett. {\bf 79},
5082 (1997).
\bibitem{nanexp}M. Bockrath, {\it et al.}, Nature {\bf 397}, 598 (1999); 
Z. Yao, H. Postma, L. Balents, C. Dekker, Nature {\bf 402}, 273 (1999);
H. Postma, M. de Jonge, Z. Yao, C. Dekker, cond-mat/0009055. 
\bibitem{Andreev} A. F. Andreev, Zh. Eksp. Teor. Fiz. {\bf 46}, 1823
(1964) [JETP {\bf 19}, 1228 (1964)]; Zh. Eksp. Teor. Fiz. {\bf 49}, 655
(1965) [JETP {\bf 49}, 455 (1966)];
\bibitem{Keldysh} L.V. Keldysh, ZhETF {\bf 47} 1515 (1964) [Sov. Phys. 
JETP {\bf 20}, 1018 (1965)]; 
M.P.A. Fisher and W. Zwerger, Phys. Rev. B {\bf 32}, 6190 (1985).
\bibitem{FG} M. P. A. Fisher and L. I. Glazman, Mes. Elec. Transp., 
ed. by L.L. Sohn, L.P. Kouwenhoven, and G. Schon, NATO Series E, 
Vol. 345, 331 (Kluwer Academic Publishing, Dordrecht, 1997).
\bibitem{KF} C. L. Kane and M. P. A. Fisher, Phys. Rev. Lett. 
{\bf 72}, 724 (1994).
\end{references}
\end{document}